\documentclass[twocolumn,superscriptaddress,prb]{revtex4}
\pdfoutput=1
\usepackage{graphicx}
\usepackage{dcolumn}
\usepackage{bm}
\usepackage{color}%
\usepackage{amsmath}%
\usepackage{amsfonts}%
\usepackage{amssymb}
\providecommand{\U}[1]{\protect\rule{.1in}{.1in}}
\renewcommand{\deg}{$^\circ$}
\begin{document}
\title{Analogue of electromagnetically induced transparency in a terahertz metamaterial}
\author{Sher-Yi Chiam}
\affiliation{Department of Physics, Science Drive 3, National University of Singapore,Singapore 117542}
\author{Ranjan Singh}
\affiliation{School of Electrical and Computer Engineering, Oklahoma State University, Stillwater, Oklahoma 74078,
USA}
\author{Carsten Rockstuhl}
\author{Falk Lederer}
\affiliation{Institute of Condensed Matter Theory and Solid State
Optics, Friedrich-Schiller-Universt\"{a}t Jena, Jena 07743, Germany}
\author{Weili Zhang}
\affiliation{School of Electrical and Computer Engineering, Oklahoma State University, Stillwater, Oklahoma 74078,
USA}
\author{Andrew A. Bettiol}
\email{phybaa@nus.edu.sg}
\affiliation{Department of Physics, Science Drive 3, National University of Singapore,Singapore 117542}

\date{\today}

\begin{abstract}
We experimentally demonstrate at terahertz frequencies that a planar metamaterial exhibits a spectral response resembling electromagnetically induced transparency. The metamaterial unit cell consists of a split ring surrounded by another closed ring where their dimensions are such that their excitable lowest order modes have identical resonance frequencies but very different life times. Terahertz time-domain spectroscopy verifies that the interference of these two resonances results in a narrow transparency window located within a broad opaque region. In contrast to previous studies this enhanced transmission is achieved by \emph{independently} exciting two resonances in which their coupling to the radiation field, and thus their linewidth, differs strongly. Rigorous numerical simulations prove that the transparency window is associated with a large group index and low losses, making the design potentially useful for slow light applications. This experiment opens an avenue to explore quantum mechanical phenomena using localized resonances in metallic structures.
\end{abstract}

\maketitle



Much emphasis in contemporary optics is on mimicking effects known from solid state physics and quantum mechanics in optical systems \cite{longhi,biagioni}. Because of the intrinsic coherence of photons emitted by a laser many effects, hardly accessible in atomic or solid state systems, can be easily verified and visualized in optics. Prominent examples include Bloch oscillations and Zener tunneling \cite{bloch,zener}. Moreover, these studies may extend the functionality of optical arrangements. Recently, optical metamaterials were put forward as a promising means in this context. A notable example is the successful application of the plasmon hybridization model \cite{03sci_prohan} to explain complex coupling behavior in vertically stacked cut wires \cite{liuam07} and split rings \cite{liu3D}. This allowed the resonances of complex metallic nanostructures to be explained on the basis of the coupling between individual plasmonic entities; just as molecular orbital theory explains transitions by linear combinations of atomic orbitals. It has been shown that they may be used to mimic the quantum phenomenon of electromagnetically induced transparency (EIT) \cite{firstmetaeit,asr}. This was first experimentally demonstrated in the radio frequency regime \cite{metaeit}, whereas a theoretical study has been performed in the optical regime \cite{plasmoneit}. Since then, additional work in this area has been reported \cite{tassinprl,tassinoe,liuam08eit,ranjanprl,09apl_papa,09prbyanno,09natmat_liu}.

EIT refers to the phenomenon where an otherwise opaque atomic medium is rendered transparent to a probe laser beam by a second, coupling beam \cite{eitmodphy,eitprl}. The presence of the coupling beam results in a transparency window of narrow spectral width in the absorption band. EIT occurs in three-level atomic systems and can be explained by destructive quantum interference between the pump and the probe beam, which are tuned to different transitions. An
alternative explanation relies on the presence of a dark superposition state when both the probe and coupling beams are turned on. In the context of metamaterials EIT is an appealing phenomenon to study because it allows us to draw specific analogies between quantum mechanical and optical systems. The metamaterial analogy to EIT can be explained as being either the result of engaging ``trapped mode" resonances \cite{asr,metaeit,09apl_papa}, or by the coupling of a so called ``bright" and a ``dark" eigenmode \cite{plasmoneit}. In developing a strong analogy to EIT a significant difference in the quality factors (or linewidth) of the two resonances involved is required. The bright eigenmode must exhibit a strong coupling to the radiation field (large linewidth, low quality factor) whereas the dark mode should only weakly couple to this field (narrow linewidth, large quality factor) \cite{eitmodphy}. However, it has to be stressed that in all experimental implementations thus far \cite{metaeit,liuam08eit,ranjanprl,09apl_papa} the spectral properties of the dark mode could not be accessed independently since it was exclusively excited by virtue of its coupling to the bright mode. This indirect excitation mechanism weakens the analogy to EIT since the dark mode should actually be excitable regardless of its weak coupling to the radiation field. In order to draw a meaningful analogy between the quantum phenomenon of EIT and its counterpart in metamaterials, a clear and unanimous relation between terms coined in quantum physics and optics is required.

In this work we are going to solve this problem. We experimentally study a planar metamaterial with a modified split ring resonator (SRR) design. This consists of a metallic split ring enclosed within a larger, closed metallic ring. The design ensures that the fundamental eigenmodes of the inner and the outer ring, assigned respectively as the dark and the bright mode, have identical frequencies but strongly deviating linewidths. Terahertz Time-Domain Spectroscopy (THz-TDS) confirms that this modified structure mimics EIT. Complementary rigorous tools were used to study the structure theoretically. Again it is worth mentioning that in this experiment the dark eigenmode of the inner ring remains excitable even in the absence of the closed outer ring. This contribution shall stipulate further research in which localized resonances in metallic structures can be used to map problems from solid state physics, quantum mechanics or quantum optics to classical optical analogies; opening a wide range of new experiments.

\begin{figure}[ptbh]
\begin{center}
\includegraphics[height=3cm]{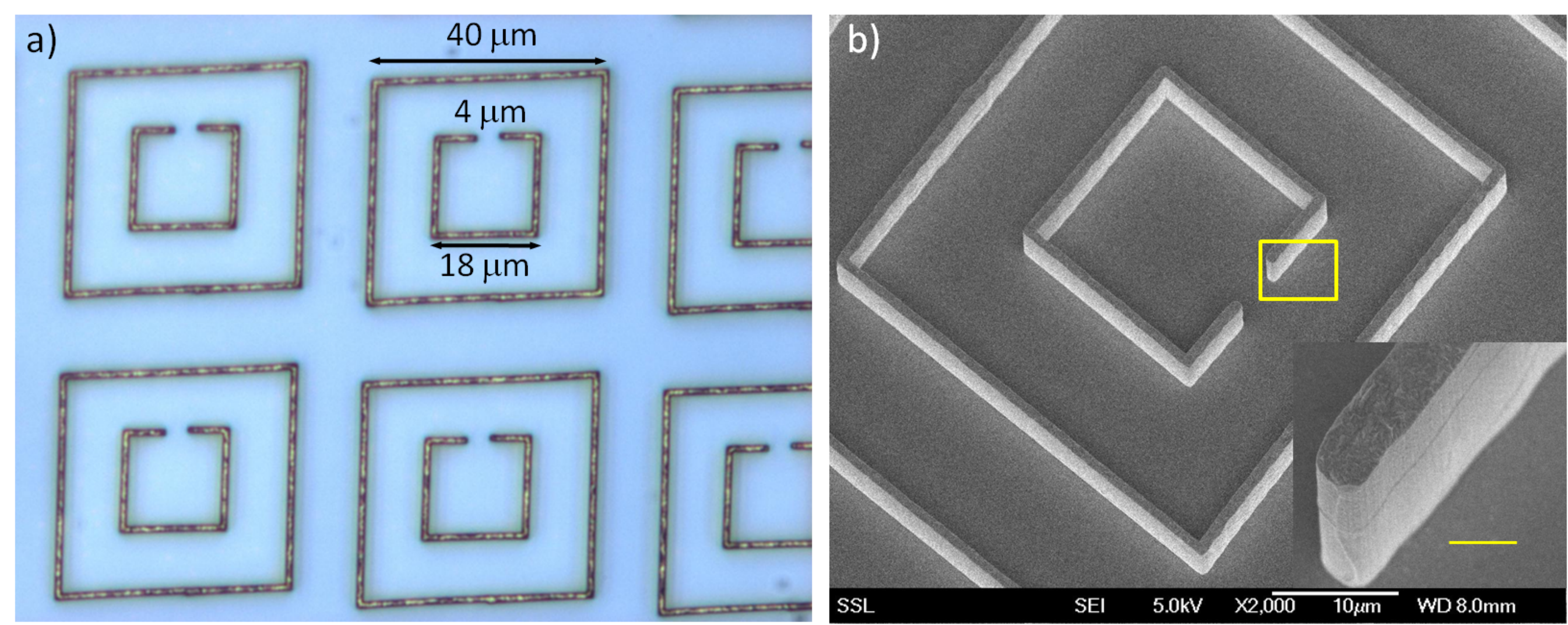}
\end{center}
\par
\vspace{-0.5cm} \caption[pics]{(Color Online) Optical (a) and scanning electron (b) micrographs of the sample fabricated by proton beam writing. The insert in (b) shows details of the region marked in the main panel (scale bar 1 $\mu $m). The width of the SRR arms is about 800 nm, and the height is over 4 $\mu$m. }%
\label{pics}%
\end{figure}

Figure \ref{pics} shows optical and scanning electron micrographs of the fabricated sample. Gold structures on a silicon substrate were fabricated using the Proton Beam Writing (PBW) technique \cite{pbwmat2day,pbwijn}. The PBW lithography process is based on patterning thick resist layers using a focused proton beam. The large mass of the protons (relative to electrons) allows the protons to maintain straight tracks through many microns of resist. An additional electroplating step defines metallic structures with high aspect ratio.

\begin{figure}[ptbh]
\begin{center}
\includegraphics[height=5cm]{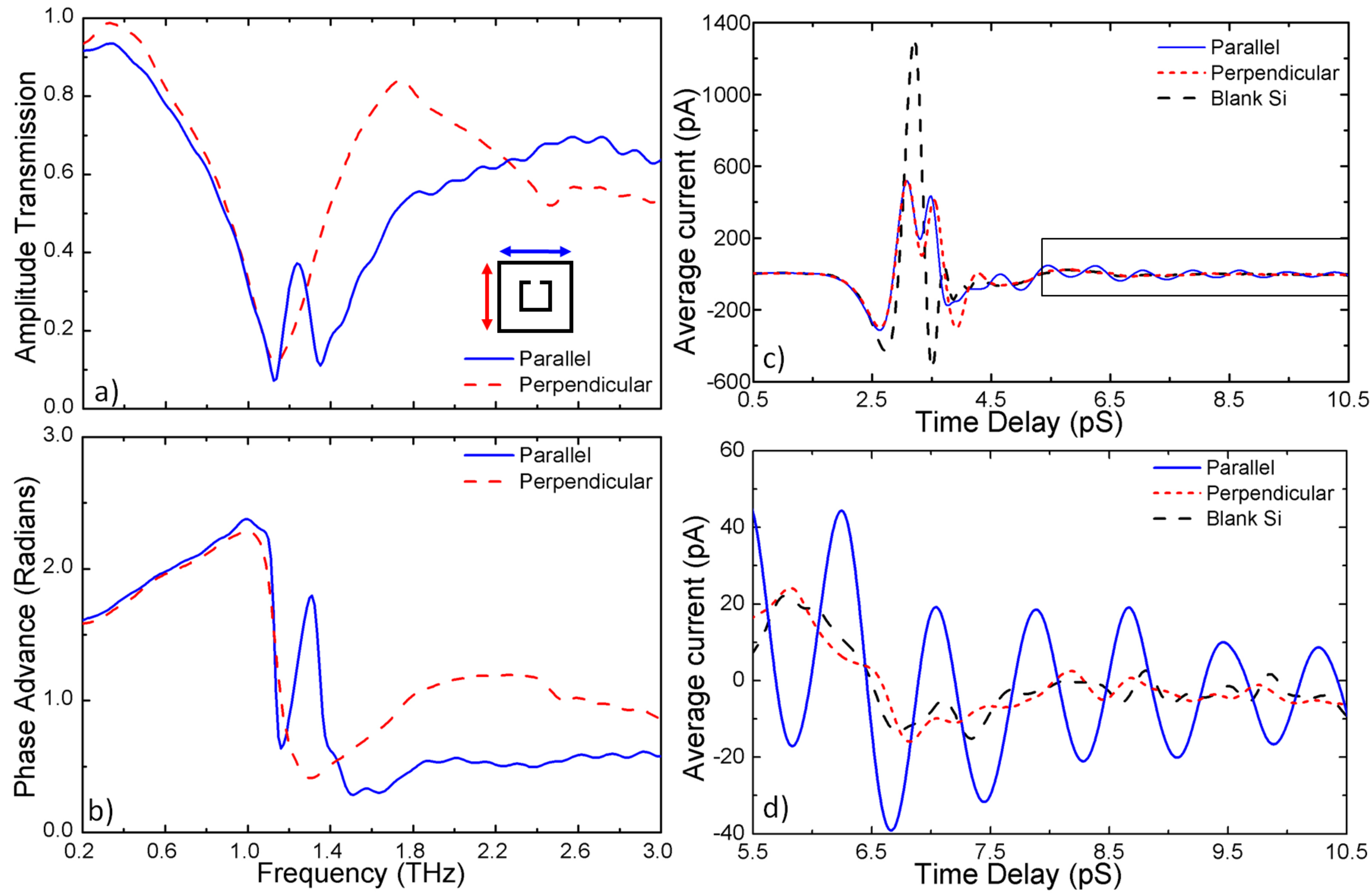}
\end{center}
\par
\vspace{-0.5cm} \caption[exspec]{(Color Online) Measured amplitude transmission (a) and phase advance (b) for the fabricated sample as a function of frequency and for two polarization states of the illumination. Inserts show the orientation of the $\mathbf{E}$ field for parallel (blue, solid) and perpendicular (red,dashed) polarization. The measured data in time domain are shown in (c) where the slowly decaying oscillations at 1.25 THz under parallel polarization in the marked area are zoomed in (d).}%
\label{exspec}%
\end{figure}

Figure \ref{exspec} shows the experimentally measured THz spectra of the sample. The spectra were collected with the beam normal to the sample plane and two different polarizations are shown. Sample characterization was carried out by the use of a photoconductive switch-based THz-TDS system, in which four parabolic mirrors are arranged in an $8-f$ confocal geometry \cite{utopsetup}. The spectra were measured using the blank Si wafer as a reference.

It is evident from Fig. \ref{exspec}(a) that the transmission spectrum shows a broad dip centered at about 1.25 THz for
polarization perpendicular to the gap of the inner SRR (i.e. perpendicular polarization). When the electric field is orientated parallel to the gap of the inner SRR (i.e. parallel polarization), a small transparency window at 1.25 THz opens where the maximum amplitude transmission exceeds 0.3. Complementarily, for perpendicular polarization, the phase data [Fig. \ref{exspec}(b)] shows a region of anomalous dispersion as typically associated with a resonance. For parallel polarization a region of strong normal dispersion appears within the transparency window. Time domain data shows a slowly decaying current oscillation with a carrier frequency of about 1.25 THz [Fig. \ref{exspec}(d)] matching exactly the central frequency of the transparency window. The temporal shift of the pulse is a signature of the group velocity reduction at 1.25 THz.

\begin{figure}[ptbh]
\begin{center}
\includegraphics[height=9cm]{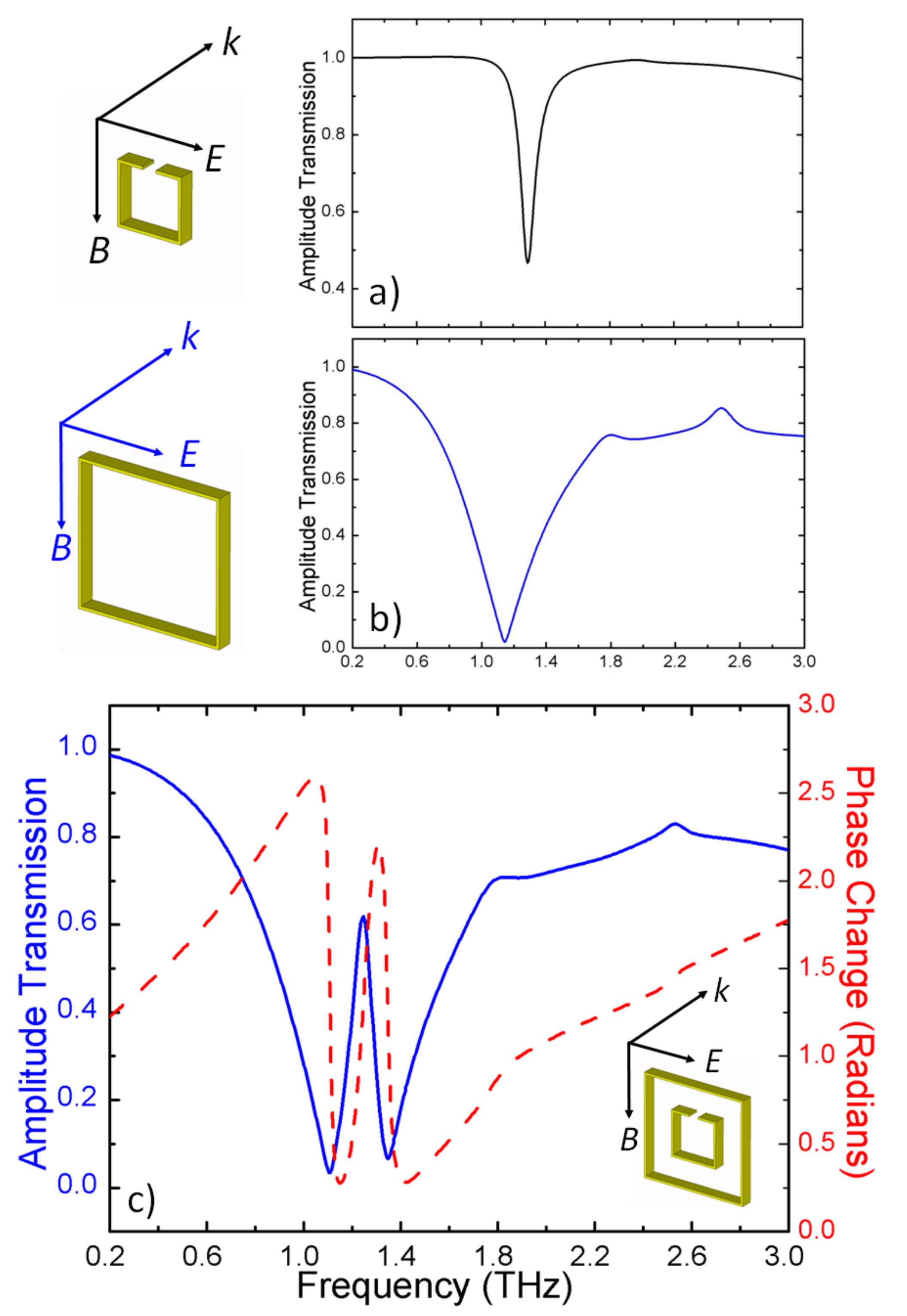}\newline
\end{center}
\par
\vspace{-0.5cm} \caption[simspec]{(Color Online) Simulated
transmission spectra of the double ring structure and its isolated
constituents;  inner split ring (a), closed outer ring (b),
 and double ring (c)  (blue solid curve, left
axis). The red dashed curve (right axis) in (c)  shows additionally the phase advance. }%
\label{simspec}%
\end{figure}

Spectra simulated using the commercial software package Microwave Studio$^{TM}$ are shown in Fig. \ref{simspec}. We show the simulated spectra of the inner split ring alone (a) and the larger closed ring alone (b); both under normal incidence with parallel polarization. We see that both structures have independently excitable resonances
leading to transmission dips and centered around the same frequency of about 1.2 THz. The Q factors (defined as the ratio of resonance frequency to the bandwidth at 3dB) of the two resonances differ by about an order of magnitude. For the split ring, the transmission dip is sharp and has a Q factor of about 11.3. The transmission dip of the closed outer ring has a Q factor of 1.25. When combined into a single structure with the split ring centered within the closed
outer ring, the interference of the resonances leads to a transparency window, as verified by the measured spectra. From Fig. \ref{simspec}(c) it is evident that the simulated transmission spectra (left axis) as well as phase data (right axis) match the experimental data well. The simulated results were verified with an independent code based on the Fourier Modal Method \cite{li}. Results from both simulations were in excellent agreement.

The resonances we observe in Fig. \ref{simspec}(a) and (b) apply to the lowest order eigenmodes of the respective structure. For the inner split ring the observed eigenmode at 1.25 THz is the so called $LC$ resonance \cite{linden}, also interpreted as the fundamental, odd eigenmode \cite{reinterpret}. Simulations reveal that the eigenmode current of the inner split ring is circular with no direct electric dipole moment. It is therefore regarded as a dark mode that
couples only weakly to the radiation field. The closed outer ring does not support the odd eigenmodes due to its symmetry. Its lowest order eigenmode is thus even and can be regarded as an electric dipole resonance similar to that of cut wires \cite{koschny}. At resonance, the current oscillates symmetrically in the two sides arms of the closed ring parallel to the electric field, resulting in a significant electric dipole moment. This resonance thus couples
strongly to the radiation field and is considered the bright mode. It is evident that the interference of these two resonances, which both lead individually to transmission dips, results in a transparency window. This EIT-like effect is observed only under parallel polarization, when both resonances can be excited. The odd eigenmode of the inner split ring cannot be excited under perpendicular polarization due to symmetry constraints \cite{gay,katsa_elect}.

From the transmission and phase data it can be recognized that strong dispersion, leading to a huge group index, occurs in the transparency window. This indicates that a light pulse with a center frequency situated in the transparency window will be considerably slowed down upon traversing the metamaterial. This behavior is also characteristic of the EIT phenomenon in atomic media. When EIT occurs in a three level system, anomalous dispersion normally observed for a two level system, is modified to a very steep normal dispersion at the transparency window, resulting in a drastic
reduction of the speed of light \cite{eitslowlight}. The controlled reduction of the speed of light attracts much interest due to its potential for practical applications, such as regenerators for optical communication and light storage. However, the quantum EIT phenomenon is sensitive to broadening by atomic motion. Therefore, the setups must typically be cooled to liquid helium temperatures, making its application to practical systems difficult. The mimicking of EIT in metamaterials is thus an attractive means to develop the building blocks of systems for slow light applications.

\begin{figure}[ptbh]
\begin{center}
\includegraphics[height=4.5cm]{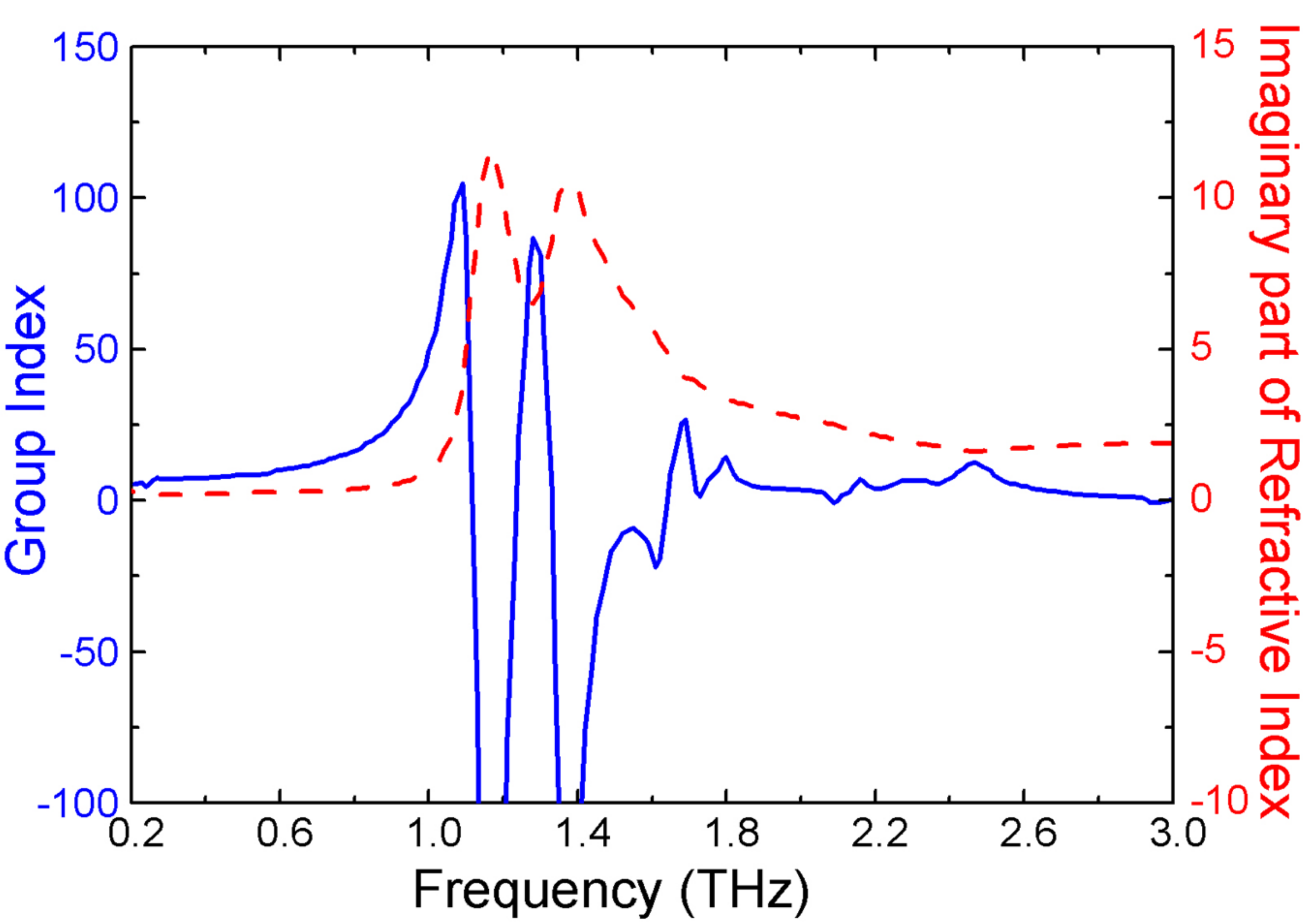}
\end{center}
\par
\vspace{-0.5cm} \caption[index]{(Color online) Group index (blue
solid curve, left axis) and imaginary part of the refractive index
(red dashed curve, right
axis) as retrieved from simulated data.}%
\label{index}%
\end{figure}

Figure \ref{index} shows the group index and the imaginary part of the refractive index retrieved from the numerical simulations. At the center of the transparency window at 1.25 THz the group index attains a peak value exceeding 75, while the imaginary part of the refractive index is as small as 5. This further verifies our experimental observation from the time domain data [Fig. \ref{exspec}(d)]. For practical applications this coincidence is of major importance and underlines the potential of this simple planar metamaterial  for slow light applications.

For the purpose of developing analogies between quantum mechanical and optical systems, the advantage of our present design is that the two eigenmodes can be independently excited and their Q factors are determined by design parameters of the ring rather by the mutual coupling strength. Our assignment of the bright and dark modes is purely based  on their individually determined Q-factors. This contrasts with some previous schemes, where the dark mode was not
directly excitable \cite{plasmoneit,liuam08eit,ranjanprl,tassinprl}. A pivotal example of such a design consists of a pair of split rings, where one of which is rotated 90\deg\space with respect to the other \cite{liuam08eit,ranjanprl}. The bright entity is the split ring whose gap is orientated parallel to the incident electric field. A corresponding resonance can be excited in the second, dark spilt ring only due to the presence of the first ring, whose presence breaks the symmetry. In our current design shifting the inner ring off-center breaks the symmetry too. Simulations show that this makes the dark mode indirectly excitable under perpendicular polarization and a finite transparency window appears. However, an individual access to estimate the linewidth of the dark mode is likewise impossible in this case.

In metamaterials, where only the bright eigenmode is directly excitable, it is difficult to draw a direct analogy to  the quantum phenomenon of EIT. Alzar \emph{et al}. presented a classical analog to EIT consisting of two coupled harmonic oscillators where only one oscillator is harmonically driven \cite{alzar}. The driving force attains the role of the probe beam and the coupling of the oscillators mimics the  effect of the pump beam in EIT. Similarly, for metamaterials, where only the bright mode is ``driven" by the radiation field, coupling between the bright and dark modes represents the effect of the pump beam. However, such a model does not permit a precise and consistent mapping of terms between the quantum and optical phenomenon. For EIT in a three-level atomic system, the pump beam is tuned to the transition between a metastable level ($|3\rangle$) and an excited level ($|2\rangle$), while the probe beam is tuned to the transition between the ground state ($|1\rangle$) and the excited state ($|2\rangle$). Destructive
quantum interference between different pathways then leads to a transparency window. A robust, metastable level, to which the dark eigenmode in metamaterials is naturally analogous, is necessary for EIT to occur in an atomic medium. In our current model, the dark eigenmode of the inner ring may be mapped to the $|3\rangle$-$|2\rangle$ transition and the bright eigenmode to the $|1\rangle $-$|2\rangle$ transition. The classical interference of the electromagnetic fields is mapped to the quantum interference in EIT. As the frequencies of the bright and dark eigenmodes are equivalent, metastable and ground level can be regarded to be degenerated and probe and pump beam have identical frequencies.
Changing the polarization of the exciting beam in our optical system has the effect of turning the pump beam on or off in EIT.

In summary, we have studied a metamaterial where two independently excitable resonances with strongly deviating Q factors interfere. The main result of this work is that we have experimentally verified that EIT-like behavior can be achieved in this way, instead of coupling a bright mode to an otherwise inaccessible dark mode. In doing so we are able to draw a more direct analogy between the quantum phenomenon and its classical counterpart. We stress, however, that there is no real classical equivalent to the quantum interference that produces EIT in an atomic medium; and that this
work merely attempts to mimic the quantum phenomenon more precisely.

The work was funded partially by the National University of Singapore grant NUS R144 000 204 646, the U.S. National Science Foundation, as well as by the German Federal Ministry of Education and Research (Metamat) and the State of Thuringia (ProExzellenz Mema). Some computations utilized the IBM p690 cluster JUMP of the Forschungszentrum J\"{u}lich, Germany. We thank Dr G S Agarwal for fruitful discussions.


\end{document}